\documentclass[10pt]{article}
\usepackage{fullpage}
\usepackage{amsmath}
\usepackage{amssymb}
\usepackage{amsfonts}
\usepackage[dvips]{epsfig}
\usepackage{color}
\usepackage{cite}
\graphicspath{{figures/}}

\begin{document}

\begin{center}

\LARGE{ {\bf Modulating quantum fluctuations of scattered lights in disordered media via wavefront shaping
}}
\end{center}
\begin{center}
\vspace{10mm} \large

{\bf Dong Li}$^{1,2,}$\footnote{Corresponding author. E--mail: lidong@mtrc.ac.cn} and {\bf Yao Yao}$^{1,2,}$\footnote{Corresponding author. E--mail: yaoyao@mtrc.ac.cn}

\vspace{2mm}

$^1$\emph{Microsystems and Terahertz Research Center, China Academy of Engineering Physics, Chengdu Sichuan 610200, China}

 \vspace{2mm}

 $^2$\emph{Institute of Electronic Engineering, China Academy of Engineering Physics, Mianyang Sichuan 621999, China}

\vspace{5mm}
\normalsize

\end{center}

\begin{center}
\vspace{15mm} {\bf Abstract}
\end{center}

After multiple scattering of quadrature-squeezed lights in a disordered medium, the quadrature amplitudes of the scattered modes present an excess noise above the shot-noise level [Opt. Expr. 14, 6919 (2006)]. A natural question is raised whether there exists a method of suppressing the quadrature fluctuation of the output mode. The answer is affirmative. In this work, we prove that wavefront shaping is a promising method to reduce the quantum noise of quadrature amplitudes of the scattered modes. This reduction is owing to the destructive interference of quantum noise. Specifically, when the single-mode squeezed states are considered as inputs, the quantum fluctuation can always be reduced, even below the shot-noise level. These results may have potential applications in quantum information processing, for instance, sub-wavelength imaging using the scattering superlens with squeezed-state sources.

\vspace{5mm}


\section{Introduction}

The studies of nonclassical lights illuminating on a disordered medium have received increasing attentions in recent years \cite{b1998,smolka2009,peeters2010,Beenakker2000,lodahl2005a,lodahl2005b,patra1999,patra2000,two2002,wiersma2013,lahini2010,gilead2015,defienne2016,leonetti2013,starshynov2016,an2018,walschaers2016,zhang2018,vellekoop2010np,rotter2017,xu2017a,xu2017b}, which is due to the fact that this quantum optical system has significant implications for quantum information processing, including Heisenberg-limit resolution imaging \cite{hong2017,hong2018}, programmable quantum optical circuit \cite{wolterink2016,huisman2014,defienne2014}, quantum communication \cite{b2017}, and quantum optical authentication \cite{goorden2014,nikolopoulos2017,yao2016,li2017}. 

As a typical nonclassical state, the squeezed state is of importance because it can achieve lower quantum noise than the quantum fluctuation of coherent state (\textit{or} equivalently the shot noise) \cite{walls1983,walls2007,barnett2002,lvovsky2015}. As a consequence, the squeezed state can enhance signal-to-noise ratio \cite{caves81,yurke86,xiao1987precision} and has been utilized in different applications ranging from quantum imaging \cite{beskrovnyy2005,sokolov2004} to gravitational wave detection \cite{aasi2013,barsotti2018,mehmet2018}.

\begin{figure*}[h]
\begin{center}
\includegraphics[width=.90\textwidth]{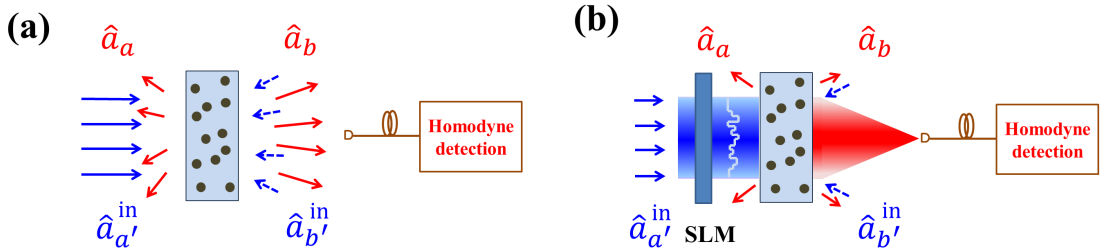} {}
\end{center}
\caption{Quadrature fluctuation detections of beams transmitted through a disordered medium (a) in the absence of WFS, (b) in the presence of WFS. $\hat{a}_{a'}^{\rm{in}}$ ($\hat{a}_{b'}^{\rm{in}}$) represents the annihilation operator of the input mode and $\hat{a}_{a}$ ($\hat{a}_{b}$) the output mode. The disordered medium, with transport mean free path $l$, thickness $L$, and number of transmission channels $N$, comprises randomly distributed small particles for light scattering. When the beams are injected, without WFS in (a), the medium separates the lights into different optical channels randomly. As a consequence, the output is in a speckle pattern. In (b), with WFS, the medium couples the beams into the desired optical paths. Hence the output presents an ordered pattern. The WFS, performed by a spatial light modulator (SLM) in (b), controls the phase of incident light. In the scheme, the focus is on the quadrature of the scattered mode, monitored by homodyne detection. }
\label{fig1}
\end{figure*}

From the perspective of quantum theory of light, the multiple scattering of squeezed states in disordered media [Fig. \ref{fig1}(a)] is a question of general interest, which has been explored from different aspects recently. For example, Ott \textit{et al.} \cite{ott2010} investigated the pairwise entanglement of scattered beams in 2010. It was found that the entanglement can be induced by multiple scattering of the squeezed state. Recently, our previous work \cite{li2019} examined the statistical distribution of quantum correlation of the scattered modes. Lodahl \cite{lodahl2006b} studied the quantum fluctuation of scattered modes and found that the averaged quantum fluctuation of quadrature amplitudes of the scattered modes is always larger than the shot noise. 

Intriguingly, it is noteworthy that in Ref. \cite{lodahl2006b} the input is a single-mode squeezed state with a sub-shot noise whereas the output exhibits an excess noise [i.e. noise above the shot-noise level, (SNL)]. In other words, after multiple scattering, the mean quantum noise of the scattered modes becomes larger than the incident one. Since the large quantum noise is detrimental for precision detection, we wonder whether there exists a method to suppress the output quantum noise.

Actually, wavefront shaping (WFS) is an emerging technology for optical focusing and imaging through disordered media \cite{vellekoop2007,vellekoop2008,popoff2010,mosk2012}, by controlling the incident wavefront, which paves a way for manipulating the speckle pattern in a desired manner. In general, WFS can be performed by spatial light modulator (SLM) in experiments as depicted in Fig. \ref{fig1}(b). The SLM acting as a reprogrammable matrix of pixels imprints expected phase values, $\phi_n$, on the coherent wavefront. In recent decades, it has been extensively utilized in numerous optical applications, such as, quantum simulator \cite{prl2019}, quantum data locking \cite{lum2016}, and super-resolution imaging \cite{putten2011,park2014,jang2018,chen2018}.

In this manuscript, we propose a scheme to modulate \textit{or} reduce the quantum fluctuation of scattered modes of a disordered medium using WFS. Two kinds of input states are considered: single- and two-mode squeezed states. We investigate the quantum noise of scattered beams in the presence of WFS. For comparison, the quantum noise in the absence of WFS is also studied. It is found that WFS can effectively reduce the quantum fluctuation for both squeezed-state inputs. In addition, the effect of disorder strength on the reduction of quantum noise is also discussed. On top of that, an intuitive explanation is given for quantum-noise reduction via WFS.

This manuscript is organized as follows: in Sec. 2, it briefly describes the model of propagation of quantized lights through a disordered medium. Sec. 3 shows how WFS reduces the quantum fluctuation of scattered modes with squeezed states as inputs. In Sec. 4, it compares the multiple-port disordered medium with the two-port beam splitter and explains the reduction of quantum noise via WFS. Sec. 5 is devoted to the conclusion of the main results. Finally, Sec. Appendix provides the derivations in detail.

\section{Theoretical model}

Fig. \ref{fig1}(a) describes the propagation of quantized lights through a disordered medium. The medium comprises randomly distributed small particles for light scattering. To characterize the medium, two primary factors are introduced: transport mean free path $l$ and thickness $L$. If $l \ll L$, the multiple scattering events would occur and result in a speckle pattern \cite{beenakker1997}. Hereafter we define $s \equiv L/l$ which determines the degree of disorder. 

\subsection{Propagation of quantized lights through a disordered medium}

After multiple scattering, the scattered mode $b$ can be written as
\begin{align}
\hat{a}_{b} = \sum_{a'}{t_{a'b}  \hat{a}_{a'}^{\rm{in}}} + \sum_{b'}{r_{b'b}  \hat{a}_{b'}^{\rm{in}}},
\label{creation}
\end{align}
where the operators $\hat{a}^{\rm{in}}_{a'}$ and $\hat{a}^{\rm{in}}_{b'}$ describe the quantum state of all open input channels and the transmission and reflection coefficients $t_{a' b}$ and $r_{b'b}$ are complex Gaussian random variables \cite{beenakker1997,rossum1999,goodman2015}. Hence $t_{a'b} = \sqrt{T_{a'b}} e^{i\phi_{a'b}}$ and $r_{b'b} = \sqrt{R_{b'b}} e^{i\phi_{b'b}}$ where $\phi_{a'b}$ ($\phi_{b'b}$) is uniformly distributed in the interval [$0,2\pi$] while $T_{a'b}$ and $R_{b'b}$ are the variables of Gaussian distribution (Particularly, $\sqrt{T_{a'b}}$ and $\sqrt{R_{b'b}}$ obeys Rayleigh distribution \cite{goodman2015} which will be used in the later section). In addition, the ensemble-averaged transmission and reflection coefficients are given by $\overline{T_{a'b}} = 1/(Ns)$ and $\overline{R_{b'b}} = (1-1/s)/N$ \cite{rossum1999,lodahl2006b}, where $N$ denotes the number of transmission channels and the overline means the averaged value over ensembles. It is shown that with the increase of disorder strength $s$, the averaged transmission coefficient $\overline{{T_{a'b}}}$ decreases. It is worthy pointing out that Eq. (\ref{creation}) quantifies the very general coupling between the input modes and the output modes. The specific characteristics of the multiple scattering disordered medium are represented by the reflection and transmission coefficients. For instance, $t_{a'b}$ describes the connection between the output mode $b$ and the input mode $a'$. 

Mathematically, the quadrature operators are introduced as $\hat{x} = \hat{a}^{\dagger} + \hat{a}$ and $\hat{p} = i(\hat{a}^\dagger - \hat{a})$. According to Eq. (\ref{creation}), the quadrature amplitudes of the scattered mode $b$ are then found to be 
\begin{align}
\label{x0}
\hat{x}_b =& \sum_{a'}{\sqrt{T_{a'b}} [\cos \phi_{a'b} \hat{x}_{a'}^{\rm{in}} - \sin \phi_{a' b} \hat{p}_{a'}^{\rm{in}}]} \\ \nonumber
&+ \sum_{b'}{\sqrt{R_{b'b}} [\cos \phi_{b'b} \hat{x}_{b'}^{\rm{in}} - \sin \phi_{b'b} \hat{p}_{b'}^{\rm{in}}] },
\end{align}
\begin{align}
\hat{p}_b =& \sum_{a'}{\sqrt{T_{a'b}} [\cos \phi_{a'b} \hat{p}_{a'}^{\rm{in}}+ \sin \phi_{a' b} \hat{x}_{a'}^{\rm{in}}]}  \\ \nonumber
&+ \sum_{b'}{\sqrt{R_{b'b}} [\cos \phi_{b'b} \hat{p}_{b'}^{\rm{in}} + \sin \phi_{b'b} \hat{x}_{b'}^{\rm{in}}] }.
\end{align}

\subsection{Modified propagation via wavefront shaping}

The original input-output relation [Eq. (\ref{creation})] can be modified via WFS \cite{vellekoop2007} as
\begin{align}
\hat{a}_{b}^{\rm{w}} = \sum_{a'}{|t_{a'b }|  \hat{a}_{a'}^{\rm{in}}} + \sum_{b'}{r_{b'b }  \hat{a}_{b'}^{\rm{in}}},
\label{creation2}
\end{align}
where the superscript $w$ denotes WFS. In the modified relation, $|t_{a'b }| $ takes the place of the complex transmission coefficient $t_{a'b}$ in Eq. (\ref{creation}), which results from the fact that the phase modulator exactly compensates the phase retardation in the disordered medium for each transmission channel, i.e. $\phi_{n} = -\phi_{a'b}$.

According to Eq. (\ref{creation2}), the quadrature operators of scattered modes in the presence of WFS correspondingly arrive at
\begin{align}
\label{xp1}
\hat{x}_b^{\rm{w}} = \sum_{a'}{\sqrt{T_{a'b}} \hat{x}_{a'}^{\rm{in}} } + \sum_{b'}{\sqrt{R_{b'b}} [\cos \phi_{b'b} \hat{x}_{b'}^{\rm{in}} - \sin \phi_{b'b} \hat{p}_{b'}^{\rm{in}}] },
\end{align}
and
\begin{align}
\label{xp2}
\hat{p}_b^{\rm{w}} = \sum_{a'}{\sqrt{T_{a'b}} \hat{p}_{a'}^{\rm{in}} } + \sum_{b'}{\sqrt{R_{b'b}} [\cos \phi_{b'b} \hat{p}_{b'}^{\rm{in}} + \sin \phi_{b'b} \hat{x}_{b'}^{\rm{in}}] }.
\end{align}

Our proposal can be completely realized in experiments under the current condition in laboratory nowadays, since this setting of phase modulation has been intensively investigated in theory and experiments over recent decades \cite{vellekoop2007,vellekoop2008,popoff2010tm,yoon2015,lerosey2007,tay2014,wang2015,ojambati2016,mounaix2016,fang2017,peng2018,stern2019}. However, different from previous works concentrating mainly on the enhanced intensity of the focused mode \cite{vellekoop2007,vellekoop2008,thompson2016,osnabrugge2019}, our work will focus on the modified quantum fluctuation of the scattered mode.

\section{Variance of quadrature of the scattered modes}

The variance of operator $\hat{O}$ is defined as
\begin{align}
\label{var20}
\langle (\Delta \hat{O})^2 \rangle \equiv \langle \hat{O}^2 \rangle - \langle \hat{O} \rangle^2,
\end{align}
where $\hat{O} = \hat{x}_b^{\rm{w}},\hat{p}_b^{\rm{w}}$. That is to say, to obtain the variances, it requires to compute $\langle \hat{x}_b^{\rm{w}} \rangle, \langle \hat{p}_b^{\rm{w}} \rangle, \langle (\hat{x}_b^{\rm{w}})^2 \rangle$, and $\langle (\hat{p}_b^{\rm{w}})^2 \rangle$.

Assume that the input modes of the right-hand side of the disordered medium are all the vacuum states (namely $\langle \hat{x}_{b'}\rangle = \langle \hat{p}_{b'}\rangle = 0$). According to Eqs. (\ref{xp1}) and (\ref{xp2}), the expectation values of $\hat{x}_b^{\rm{w}}$ and $\hat{p}_b^{\rm{w}}$ can be rewritten as
\begin{align}
\label{var2a}
\langle \hat{x}_b^{\rm{w}} \rangle = \sum_{a'}{\sqrt{T_{a'b}}  \langle \hat{x}_{a'}^{\rm{in}} \rangle}, \\ \nonumber
\langle \hat{p}_b^{\rm{w}} \rangle = \sum_{a'}{\sqrt{T_{a'b}}  \langle \hat{p}_{a'}^{\rm{in}} \rangle}.
\end{align}
Note that $\langle \hat{x}_b^{\rm{w}} \rangle$ ($\langle \hat{p}_b^{\rm{w}} \rangle$) is only related to the transmitted modes $\langle \hat{x}_{a'}^{\rm{in}} \rangle$ ($\langle \hat{p}_{a'}^{\rm{in}} \rangle$) due to the vacuum states for all reflected ones. 

From Eqs. (\ref{xp1}) and (\ref{xp2}), the mean values of $(\hat{x}_b^{\rm{w}})^2$ and $(\hat{p}_b^{\rm{w}})^2$ are found to be
\begin{align}
\label{var2b}
\langle (\hat{x}_b^{\rm{w}})^2 \rangle  =&  \sum_{a'\neq a''}  \sqrt{T_{a'b} T_{a''b}} [  \langle\hat{x}_{a'}^{\rm{in}} \hat{x}_{a''}^{\rm{in}}\rangle +  \langle\hat{x}_{a''}^{\rm{in}} \hat{x}_{a'}^{\rm{in}}\rangle  ]\\ \nonumber
& +  \sum_{b'} R_{b'b}  [\cos^2 \phi_{b'b}  \langle(\hat{x}_{b'}^{\rm{in}})^2\rangle  + \sin^2 \phi_{b'b}  \langle(\hat{p}_{b'}^{\rm{in}})^2\rangle  \\ \nonumber
&- \cos \phi_{b'b} \sin \phi_{b'b} ( \langle\hat{x}_{b'}^{\rm{in}} \hat{p}_{b'}^{\rm{in}}\rangle +  \langle\hat{p}_{b'}^{\rm{in}} \hat{x}_{b'}^{\rm{in}}\rangle )  ] \\ \nonumber
&+ \sum_{a' b'}\sqrt{T_{a'b} R_{b'b}} [\cos \phi_{b'b} (\langle\hat{x}_{a'}^{\rm{in}} \hat{x}_{b'}^{\rm{in}}\rangle + \langle\hat{x}_{b'}^{\rm{in}} \hat{x}_{a'}^{\rm{in}}\rangle) \\ \nonumber
&- \sin \phi_{b'b} (\langle\hat{x}_{a'}^{\rm{in}} \hat{p}_{b'}^{\rm{in}}\rangle + \langle\hat{p}_{b'}^{\rm{in}} \hat{x}_{a'}^{\rm{in}}\rangle)] + \sum_{a'} {T_{a'b} \langle(\hat{x}_{a'}^{\rm{in}})^2 \rangle } ,\\ \nonumber
\langle (\hat{p}_b^{\rm{w}})^2 \rangle  =& \sum_{a'\neq a''}  \sqrt{T_{a'b} T_{a''b}} [  \langle\hat{p}_{a'}^{\rm{in}} \hat{p}_{a''}^{\rm{in}}\rangle +  \langle\hat{p}_{a''}^{\rm{in}} \hat{p}_{a'}^{\rm{in}}\rangle  ]\\ \nonumber
& +  \sum_{b'} R_{b'b} [ \cos^2 \phi_{b'b}  \langle(\hat{p}_{b'}^{\rm{in}})^2\rangle  + \sin^2 \phi_{b'b}  \langle(\hat{x}_{b'}^{\rm{in}})^2\rangle \\ \nonumber
& + \cos \phi_{b'b} \sin \phi_{b'b} ( \langle\hat{x}_{b'}^{\rm{in}} \hat{p}_{b'}^{\rm{in}}\rangle +  \langle\hat{p}_{b'}^{\rm{in}} \hat{x}_{b'}^{\rm{in}}\rangle ) ]  \\ \nonumber
&+ \sum_{a' b'}\sqrt{T_{a'b} R_{b'b}} [\cos \phi_{b'b} (\langle\hat{p}_{a'}^{\rm{in}} \hat{p}_{b'}^{\rm{in}}\rangle + \langle\hat{p}_{b'}^{\rm{in}} \hat{p}_{a'}^{\rm{in}}\rangle) \\ \nonumber
&+ \sin \phi_{b'b} (\langle\hat{p}_{a'}^{\rm{in}} \hat{x}_{b'}^{\rm{in}}\rangle + \langle\hat{x}_{b'}^{\rm{in}} \hat{p}_{a'}^{\rm{in}}\rangle)] + \sum_{a'} {T_{a'b} \langle(\hat{p}_{a'}^{\rm{in}})^2 \rangle },
\end{align}
where the expectation values are universal for any input states.

\subsection{Single-mode squeezed states as input}

In this paper, we will discuss two specific situations: single- and two-mode squeezed states as input. Consider single-mode squeezed states as input, $| \Psi_1^{\rm{in}} \rangle = [\hat{D}(\alpha)\hat{S}(r)  |0\rangle]^{\otimes K}$, with $K$ being the number of input modes, $\hat{D}(\alpha) = e^{\alpha \hat{a}^{\dagger} - \alpha^{\ast} \hat{a}}$ thedisplacement operator, and $\hat{S}(r) = e^{ (r /2)(\hat{a}^{\dagger 2} - \hat{a}^{2}) }$ the squeezing operator (the complex number $\alpha$ is the amplitude and the real number $r$ is the squeezing parameter).

\begin{figure}[htbp]
\begin{center}
\includegraphics[width=.40\textwidth]{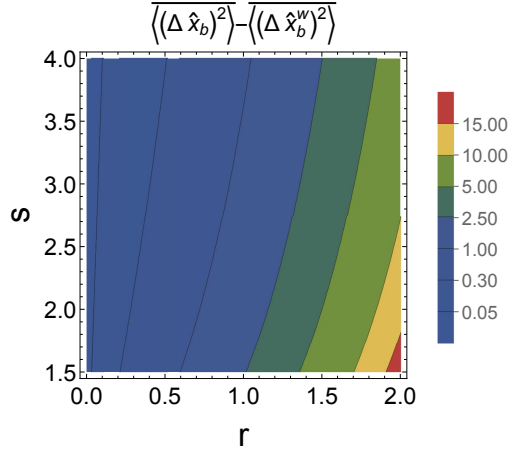} {}
\end{center}
\caption{The difference between $\overline{\langle (\Delta \hat{x}_b)^2 \rangle}$ and $\overline{\langle (\Delta \hat{x}_b^{\rm{w}})^2 \rangle}$ as a function of $r$ and $s$ with single-mode squeezed states as input ($|\Psi^{\rm{in}}_1 \rangle = [\hat{D}(\alpha)\hat{S}(r)|0\rangle]^{\otimes K}$), with the number of input modes $K$, displacement operator $\hat{D}(\alpha) = e^{\alpha \hat{a}^{\dagger} - \alpha^{\ast} \hat{a}}$, and squeezing operator $\hat{S}(r) = e^{ (r /2)(\hat{a}^{\dagger 2} - \hat{a}^{2}) }$ (complex number $\alpha$ being the amplitude and real number $r$ the squeezing parameter). It shows that the difference is always larger than zero, namely the variance without WFS is greater than the one with WFS, which elucidates that WFS can reduce the averaged variance of the scattered modes. Parameters used are: (a) $s = 2$ and (b) $r = 1$. We have set the number of input modes $K=N$ ($N$ is the number of the transmission channels of the disordered medium).}
\label{3dsqz1}
\end{figure}

Note that the quadrature operators of a single-mode squeezed state can be written as
\begin{align}
\label{quadrature}
\hat{x}_{a'}^{\rm{in}} &= x + e^{-r} \hat{x}^{\rm{v}}_{a'},\\ \nonumber
\hat{p}_{a'}^{\rm{in}} &= p + e^{r} \hat{p}^{\rm{v}}_{a'},
\end{align}
where $x$ and $p$ [$\alpha = (x + i p)/2$] are the mean values of operators $\hat{x}_{a'}^{\rm{in}}$ and $\hat{p}_{a'}^{\rm{in}}$, respectively, and the operators $\hat{x}^{\rm{v}}_{a'}$ and $\hat{p}^{\rm{v}}_{a'}$ denote the  quadratures of the vacuum states. It is easy to find that $\langle \hat{x}_{a'}^{\rm{v}} \rangle =\langle \hat{p}_{a'}^{\rm{v}} \rangle = 0$,
$\langle \hat{x}_{a'}^{\rm{in}} \rangle = x$,
$\langle \hat{p}_{a'}^{\rm{in}} \rangle = p$.
Then
$\langle (\Delta \hat{x}_{a'}^{\rm{in}})^2 \rangle = \langle (\hat{x}_{a'}^{\rm{in}})^2 \rangle - \langle \hat{x}_{a'}^{\rm{in}} \rangle^2 = e^{-2r}$, $\langle (\Delta \hat{p}_{a'}^{\rm{in}})^2 \rangle = \langle (\hat{p}_{a'}^{\rm{in}})^2 \rangle - \langle \hat{p}_{a'}^{\rm{in}} \rangle^2 = e^{2r}$,
$\langle (\Delta \hat{x}_{b'}^{\rm{in}})^2 \rangle = \langle (\Delta \hat{p}_{b'}^{\rm{in}})^2 \rangle = 1$, $\langle \hat{x}_{a'}^{\rm{in}} \hat{x}_{a''}^{\rm{in}} \rangle - \langle \hat{x}_{a'}^{\rm{in}}\rangle \langle \hat{x}_{a''}^{\rm{in}} \rangle = 0$ ($a' \neq a''$), $\langle \hat{p}_{a'}^{\rm{in}} \hat{p}_{a''}^{\rm{in}} \rangle - \langle \hat{p}_{a'}^{\rm{in}}\rangle \langle \hat{p}_{a''}^{\rm{in}} \rangle = 0$ ($a' \neq a''$). Combining all these arguments and Eqs. (\ref{var20}), (\ref{var2a}), (\ref{var2b}), one can obtain
\begin{align}
\label{sqz1xb}
\langle(\Delta \hat{x}_b^{\rm{w}})^2 \rangle &= 1 - \sum_{a'}^{K}T_{a'b} (1 - e^{-2r}), \\ \nonumber
\langle(\Delta \hat{p}_b^{\rm{w}})^2 \rangle &= 1 + \sum_{a'}^{K}T_{a'b} ( e^{2r} - 1),
\end{align}
where we have used $\sum_{a'} T_{a'b} + \sum_{b'} R_{b'b} = 1$ (see Append. \ref{derivation} in detail).

\begin{figure}[htb]
\begin{center}
\includegraphics[width=.40\textwidth]{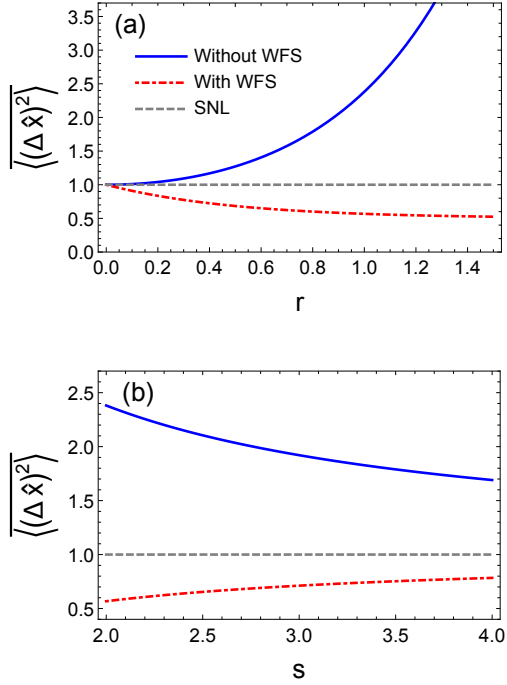} {}
\end{center}
\caption{The variances $\overline{\langle (\Delta \hat{x})^2 \rangle}$ versus (a) $r$ and (b) $s$ with single-mode squeezed states as input, $| \Psi_1^{\rm{in}} \rangle = [\hat{D}(\alpha) \hat{S}(r) |0\rangle]^{\otimes K}$. The blue-solid line denotes the quantum noise without WFS, the red-dash-dotted one the quantum noise with WFS, and the gray-dotted one the SNL. Parameters used are: (a) $s = 2$ and (b) $r = 1$. We have set the number of input modes $K=N$ ($N$ denotes the number of the transmission channels of the disordered medium). }
\label{xr}
\end{figure}

By averaging over all ensembles of disorders, Eq. (\ref{sqz1xb}) is then reduced to
\begin{align}
\label{sqz1xbm}
\overline{\langle(\Delta \hat{x}_b^{\rm{w}})^2\rangle} &= 1 - K \overline{T_{a'b}} (1 - e^{-2r}) \\ \nonumber
&= 1 - \frac{K}{Ns} (1 - e^{-2r}), \\ \nonumber
\overline{\langle(\Delta \hat{p}_b^{\rm{w}})^2\rangle} &= 1 + \frac{K}{Ns}  (e^{2r} - 1),
\end{align}
where the overline indicates the average over all ensembles ($N$ means the number of transmission channels, $K$ the number of input modes, $s$ the disorder degree, and $r$ the squeezing parameter of the input states). When $r=0$, (i.e. coherent states as input), the averaged output noise via WFS, $\overline{\langle(\Delta \hat{x}_b^{\rm{w}})^2\rangle} = \overline{\langle(\Delta \hat{p}_b^{\rm{w}})^2\rangle} = 1$, is at the shot-noise level (SNL) which is defined as the quantum fluctuation of quadrature of the coherent state ($\langle(\Delta \hat{x}_{\rm{SNL}})^2\rangle = 1$). Actually, when coherent states are injected, the scattered modes are still coherent states due to the linear splitting process in a disordered medium. Hence, the output noise is natually at the SNL. Nevertheless, when $r>0$ (i.e. squeezed states as input), it is found that the averaged output noise via WFS always reaches below the SNL ($\overline{\langle(\Delta x_b^{\rm{w}})^2\rangle}  < 1$).

For comparison, the averaged quantum fluctuation without WFS is also considered and is given by 
\begin{align}
\label{sqz1var2xxx}
\overline{\langle (\Delta \hat{x}_{b})^2 \rangle} = \overline{\langle (\Delta \hat{p}_{b})^2 \rangle} = 1+ \frac{K}{Ns} [\cosh (2r) - 1],
\end{align}
where the detailed derivation is shown in Append. \ref{appsinglemode}. When the number of input modes $K$ is reduced to one, one obtain $\overline{\langle (\Delta \hat{x}_{b})^2 \rangle} = \overline{\langle (\Delta \hat{p}_{b})^2 \rangle} = 1+ [\cosh (2r) - 1]/(Ns)$ which reproduces the result in Ref. \cite{lodahl2006b}. On the other hand, when $r>0$ (i.e. squeezed-state input), the quantum noise is always above the SNL, which indicates that the scattered mode without WFS shows a quantum noise above the SNL. The difference between the variances with WFS and without WFS is plotted in Fig. \ref{3dsqz1}. It is easily found that this difference is always positive, meaning that the variance without WFS is greater than the one with WFS, which yields that the quantum noise is degraded via WFS. 

Fig. \ref{xr}(a) compares the variances with and without WFS as a function of $r$. As shown in Fig. \ref{xr}(a), the blue-solid (red-dash-dotted, gray-dotted) line represents the variance without WFS (with WFS, SNL). It can be seen that the variance without WFS is always above the SNL whereas the one with WFS is below the SNL. In other words, the WFS can suppress the quantum noise, even below the SNL. With the increase of $r$, the variance without WFS decreases whereas the one with WFS increases. Similarly, in Fig. \ref{xr}(b), the variances of scattered modes versus $s$ are plotted. It is shown that as $s$ increases the variance without WFS decreases. On the contrary, with the increase of $s$, the variance with WFS increases.

\subsection{Two-mode squeezed states as input}

\begin{figure}[htb]
\begin{center}
\includegraphics[width=.40\textwidth]{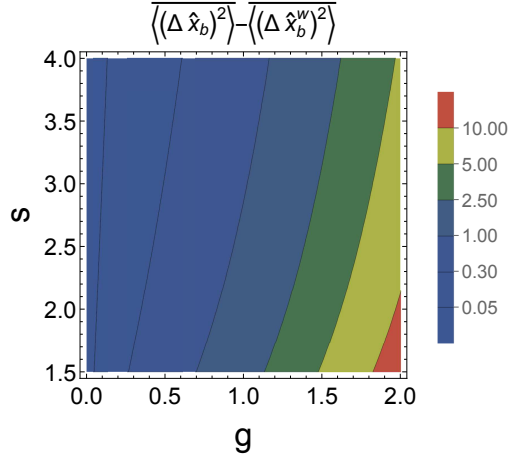} {}
\end{center}
\caption{The difference between $\overline{\langle (\Delta \hat{x}_b)^2 \rangle}$ and $\overline{\langle (\Delta \hat{x}_b^{\rm{w}})^2 \rangle}$ as a function of $g$ and $s$ with two-mode squeezed states as input, $| \Psi_2^{\rm{in}} \rangle = \{\hat{D}_{A}(\alpha)  \hat{D}_{B}(\beta) \hat{S}_{AB}(\zeta) |0\rangle_A |0\rangle_B \}^{\otimes K/2}$ with displacement operators $\hat{D}_A(\alpha) = e^{\alpha \hat{a}_A^{\dagger} - \alpha^{\ast} \hat{a}_A}$, $\hat{D}_B(\beta) = e^{\beta \hat{a}_B^{\dagger} - \beta^{\ast} \hat{a}_B}$, two-mode squeezing operator $\hat{S}_{AB}(g) = e^{(-\zeta\hat{a}_{A}^{\dagger}\hat{a}_{B}^{\dagger} +\zeta^{\ast} \hat{a}_{A}\hat{a}_{B}) }$ ($\zeta = g e^{i \phi_{g}}$, real number $g$ denoting the squeezing parameter and real number $\phi_{g}$ the squeezing angle), and the number of input modes $K$. It shows that the difference is always greater than zero, which elucidates that the WFS can reduce the averaged variance of quadratures of scattered modes. Parameters used are (a) $s = 2$, (b) $g = 0.6$. We have set the number of input modes $K=N$ ($N$ is the number of transmission channels of the disordered medium).}
\label{3dsqz2}
\end{figure}

\begin{figure}[htb]
\begin{center}
\includegraphics[width=.40\textwidth]{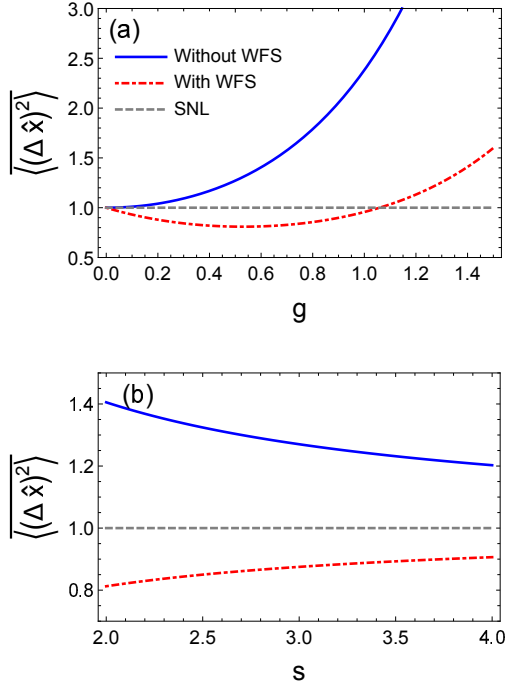} {}
\end{center}
\caption{The variances $\overline{\langle (\Delta \hat{x})^2 \rangle}$ versus (a) $g$ and (b) $s$ with two-mode squeezed states as input, $| \Psi_2^{\rm{in}} \rangle = \{\hat{D}_{A}(\alpha) \hat{D}_{B}(\beta) \hat{S}_{AB}(\zeta) |0\rangle_A |0\rangle_B \}^{\otimes K/2}$. The blue-solid line denotes the quantum noise without WFS, the red-dash-dotted one the quantum noise with WFS, and the gray-dotted one the SNL. Parameters used are (a) $s = 2$, (b) $g = 0.6$. We have set the number of input modes $K=N$ ($N$ is the number of transmission channels of the disordered medium). }
\label{xr2}
\end{figure}

Since the two-mode squeezed state \cite{barnett2002} has a squeezed fluctuation similar to the single-mode squeezed state, we wonder whether the two-mode squeezed state  could also exhibit the quantum-noise reduction via WFS.

Consider that the input is two-mode squeezed states, $| \Psi_2^{\rm{in}} \rangle = \{\hat{D}_{A}(\alpha) \hat{D}_{B}(\beta) \hat{S}_{AB}(\zeta) |0\rangle_A |0\rangle_B \}^{\otimes K/2}$ with number of the input modes $K$ ($K$ is even), displacement operators $\hat{D}_A(\alpha) = e^{\alpha \hat{a}_A^{\dagger} - \alpha^{\ast} \hat{a}_A}$, $\hat{D}_B(\beta) = e^{\beta \hat{a}_B^{\dagger} - \beta^{\ast} \hat{a}_B}$, and two-mode squeezing operator $\hat{S}_{AB}(\zeta) = e^{(-\zeta\hat{a}_{A}^{\dagger}\hat{a}_{B}^{\dagger} +\zeta^{\ast} \hat{a}_{A}\hat{a}_{B}) }$ ($\zeta = g e^{i \phi_{g}}$, the real number $g$ being the squeezing parameter and the real number $\phi_{g}$ denoting the squeezing angle).

For simplicity, let $\beta = \alpha$. Assume that $\alpha = (x + i p)/2$, and each pair of input modes $A = a'$ and $B = a'+K/2$ ($a' = 1,2,...,K/2$) constitutes a two-mode squeezed state. Then one can obtain $\langle \hat{x}_{a'}^{\rm{in}} \rangle = x$, $\langle \hat{p}_{a'}^{\rm{in}} \rangle = p$, $\langle \hat{x}_{b'}^{\rm{in}} \rangle = 0$, $\langle \hat{p}_{b'}^{\rm{in}} \rangle = 0$. And $\langle (\Delta \hat{x}_{a'}^{\rm{in}})^2 \rangle = 2 \sinh^2 g + 1$, $\langle (\Delta \hat{p}_{a'}^{\rm{in}})^2 \rangle = 2 \sinh^2 g + 1$,
$\langle (\Delta \hat{x}_{b'}^{\rm{in}})^2 \rangle = \langle (\Delta \hat{p}_{b'}^{\rm{in}})^2 \rangle = 1$. The covariance function between $\hat{x}_{a'}^{\rm{in}}$ and $\hat{x}_{a''}^{\rm{in}}$ can be described as
${\rm{cov}}(\hat{x}_{a'}^{\rm{in}}, \hat{x}_{a''}^{\rm{in}}) = \frac{1}{2}(\langle \hat{x}_{a'}^{\rm{in}} \hat{x}_{a''}^{\rm{in}} \rangle + \langle \hat{x}_{a''}^{\rm{in}} \hat{x}_{a'}^{\rm{in}} \rangle) - \langle \hat{x}_{a'}^{\rm{in}} \rangle \langle \hat{x}_{a''}^{\rm{in}} \rangle = \delta_{a',a''-K/2} 2 \cosh g \sinh g \cos \phi_g.$ It is worthy pointing out that these covariance functions vanish when the single-mode squeezed states are considered as inputs whereas they can not be ignored in the presence of two-mode squeezed input states. This is because there exists the nonclassical correlation between the two modes in each two-mode squeezed input state.

According to Eqs. (\ref{var20}), (\ref{var2a}), and (\ref{var2b}), the variances are found to be 
\begin{align}
\langle(\Delta \hat{x}_b^{\rm{w}})^2\rangle =& 1 + \sum_{a'}^{K}{T_{a'b} 2 \sinh^2 g}\\ \nonumber
&+\sum_{a'}^{K/2}{\sqrt{T_{a'b} T_{a'+K/2, b}} (4 \cos \phi_{g} \sinh g \cosh g )},
\end{align}
and
\begin{align}
\langle(\Delta \hat{p}_b^{\rm{w}})^2\rangle =& 1 + \sum_{a'}^{K}{T_{a'b} 2 \sinh^2 g}\\ \nonumber
&-\sum_{a'}^{K/2}{\sqrt{T_{a'b} T_{a'+K/2, b}} (4 \cos \phi_{g} \sinh g \cosh g )}.
\end{align}

To minimize variance $\langle(\Delta \hat{x}_b)^2\rangle$, one can set $\cos \phi_{g} = -1$ and obtain the minimum value
\begin{align}
\label{dx200a}
\langle(\Delta \hat{x}_b^{\rm{w}})^2\rangle =& 1 + \sum_{a'}^{K}{T_{a'b} 2 \sinh^2 g} \\ \nonumber
&- \sum_{a'}^{K/2}{\sqrt{T_{a'b} T_{a'+K/2, b}} (4  \sinh g \cosh g )}.
\end{align}
Meanwhile, $\langle(\Delta \hat{p}_b^{\rm{w}})^2\rangle$ is recast as
\begin{align}
\label{dp200a}
\langle(\Delta \hat{p}_b^{\rm{w}})^2\rangle =& 1 + \sum_{a'}^{K}{T_{a'b} 2 \sinh^2 g} \\ \nonumber
&+ \sum_{a'}^{K/2}{\sqrt{T_{a'b} T_{a'+K/2, b}} (4  \sinh g \cosh g )}.
\end{align}

By averaging over all disorder ensembles, the quantum fluctuations in Eqs. (\ref{dx200a}) and (\ref{dp200a}) are worked out as
\begin{align}
\label{sqz2var2}
\overline{\langle(\Delta \hat{x}_b^{\rm{w}})^2\rangle} &= 1 + 2K\overline{T_{a'b}} \sinh g ( \sinh g - \frac{\pi}{4} \cosh g)\\ \nonumber
& = 1 + \frac{2K}{Ns} \sinh g ( \sinh g - \frac{\pi}{4} \cosh g),
\end{align}
and
\begin{align}
\overline{\langle(\Delta \hat{p}_b^{\rm{w}})^2\rangle} &=  1 + \frac{2K}{Ns} \sinh g ( \sinh g + \frac{\pi}{4} \cosh g),
\end{align}
where $\overline{\sqrt{T_{a'b} T_{a'+K/2, b}}} = \overline{T_{a'b}} \pi/4$ is used and is proven in Append. \ref{rayleigh} for simplicity. Eq. (\ref{sqz2var2}) shows that if the modulated variance reaches below the SNL, it requires that $\sinh g - (\pi/4)\cosh g  < 0$, namely $ g < g_{\star} \equiv \rm{arctanh}$$( \pi / 4) \approx 1.06 $. That is to say, when $g < g_{\star}$, the output quantum noise is below the SNL. However, when $g >g_{\star}$, it does not surpass the SNL.

As a comparison, we also consider the mean variances in the absence of WFS
\begin{align}
\label{var3a}
\overline{\langle (\Delta \hat{x}_{b})^2 \rangle} &= 1+ 2 K \overline{T_{a'b}} \sinh^2 g \\ \nonumber
&= 1+ \frac{2 K}{Ns} \sinh^2 g, 
\end{align}
and
\begin{align}
\overline{\langle (\Delta \hat{p}_{b})^2 \rangle} &= 1+ \frac{2 K}{Ns} \sinh^2 g.
\end{align}
The corresponding deviation is shown in Append. \ref{apptwomode}. From Eq. (\ref{var3a}), it is noteworthy that the output quantum noise without WFS is always above the SNL. Fig. \ref{3dsqz2} plots the difference between the variances with and without WFS. It is shown that the difference is always greater than zero which indicates that WFS can reduce the quantum fluctuation of the scattered mode.

In addition, Fig. \ref{xr2}(a) compares the variances with and without WFS as a function of $g$. As shown in Fig. \ref{xr2}(a), the blue-solid (red-dash-dotted, gray-dotted) line represents the variance without WFS (with WFS, SNL). It is easy to verify that the variance with WFS is always smaller than the one without WFS. This yields that WFS can reduce the output quantum noise. Different from the single-mode-squeezed-state input, the two-mode-squeezed-state one does not beat the SNL with WFS all the time. In fact, there exists a threshold $g_{\star} \approx 1.06$ which determines whether the reduced quantum noise reach below the SNL. When $g < g_{\star}$, the variance with WFS can surpass the SNL whereas when $g > g_{\star}$, the one with WFS can not beat the SNL. In Fig. \ref{xr2}(b), the variances with and without WFS are plotted as a function of $s$. Similar to the case of single-mode squeezed states, with $s$ increasing, the variance in the absence of WFS decreases whereas the one in the presence of WFS increases as $s$ increases.

\section{Discussion}

One may wonder why WFS can modulate the quadrature fluctuations of the scattered lights of a disordered medium. Here we will provide an intuitive interpretation of the quantum-noise reduction. Before any further explanation, let us turn attention to a simple case of the conventional balanced beam splitter (BS) which is a two-port device. For simplicity, we will take the single-mode-squeezed-state input as an example.

\subsection{Comparison with the two-port beam splitter}

\begin{figure}[htbp]
\begin{center}
\includegraphics[width=.4\textwidth]{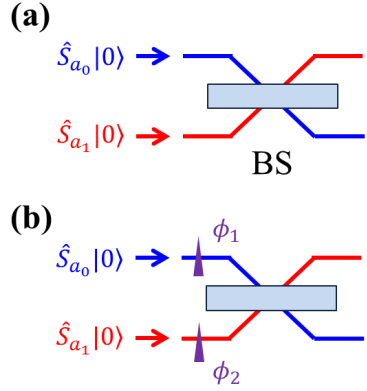} {}
\end{center}
\caption{Squeezed vacuum states propagating through a beam splitter (a) in the absence of WFS, and (b) in the presence of WFS. With a two-port beam splitter (BS), the WFS is equivalent to two phase shifters ($\phi_1$ and $\phi_2$) as depicted in (b).}
\label{bs12}
\end{figure}

For the BS, akin to the disordered medium, we consider the situation where the input is the single-mode squeezed vacuum states,
\begin{align}
|\Psi^{\rm{in}}_3\rangle = \hat{S}_{a_0}(r)|0\rangle \otimes \hat{S}_{a_1}(r)|0\rangle,
\end{align}
where $\hat{S}_{a_0}(r) = e^{(r/2) [(\hat{a}_0^{\rm{in}\dagger})^{ 2}-  (\hat{a}_0^{\rm{in} })^2]}$ ($\hat{S}_{a_1}(r) = e^{(r/2) [(\hat{a}_1^{\rm{in}\dagger})^{ 2}-  (\hat{a}_1^{\rm{in} })^2]}$) is the single-mode squeezing operator with $\hat{a}_0^{\rm{in} \dagger}$ ($\hat{a}_1^{\rm{in} \dagger}$) and $\hat{a}_0^{\rm{in}}$ ($\hat{a}_1^{\rm{in}}$) indicating the creation and annihilation operators of the input mode $0$ ($1$) and $r$ denoting the squeezing parameter. Two circumstances will be compared: (I) in the absence of WFS and (II) in the presence of WFS as illustrated in Figs. \ref{bs12}(a) and \ref{bs12}(b), respectively.

In the case I [Fig. \ref{bs12}(a)], without WFS, after passing through BS, the output state can be written as
\begin{align}
\label{outbswithoutWFS}
|\Psi^{\rm{out}}_3\rangle &= e^{i r [\hat{a}_0^{\rm{out}\dagger} \hat{a}_1^{\rm{out}\dagger} +  \hat{a}_0^{\rm{out} } \hat{a}_1^{\rm{out} }]}|0\rangle \otimes |0\rangle,
\end{align}
where $\hat{a}_0^{\rm{out} \dagger}$ ($\hat{a}_1^{\rm{out} \dagger}$) and $\hat{a}_0^{\rm{out}}$ ($\hat{a}_1^{\rm{out}}$) mean the creation and annihilation operators of the output mode $0$ ($1$). For clarity, the calculation is shown in Append. \ref{bsabsence}. As a matter of fact, the output in Eq. (\ref{outbswithoutWFS}) is the so-called two-mode squeezed vacuum state \cite{barnett2002}. After tracing over the output mode $1$, the reduced state of the output denotes the mode $0$ which is a thermal state. The corresponding variance of output mode $0$ can be worked out as
\begin{align}
\label{plus2r}
\langle (\Delta \hat{x}_{0}^{\rm{out}})^2 \rangle = 2 \sinh^2 r + 1 ,
\end{align} 
where $\hat{x}_{0}^{\rm{out}} \equiv \hat{a}_{0}^{\rm{out} \dagger} + \hat{a}_{0}^{\rm{out}}$. It is easy to check that when $r>0$, the output quantum noise of mode $0$ is always above the SNL which is similar to the case of single-mode squeezed input states in the disordered medium without WFS. 

\begin{figure*}[htb]
\begin{center}
\includegraphics[width=.75\textwidth]{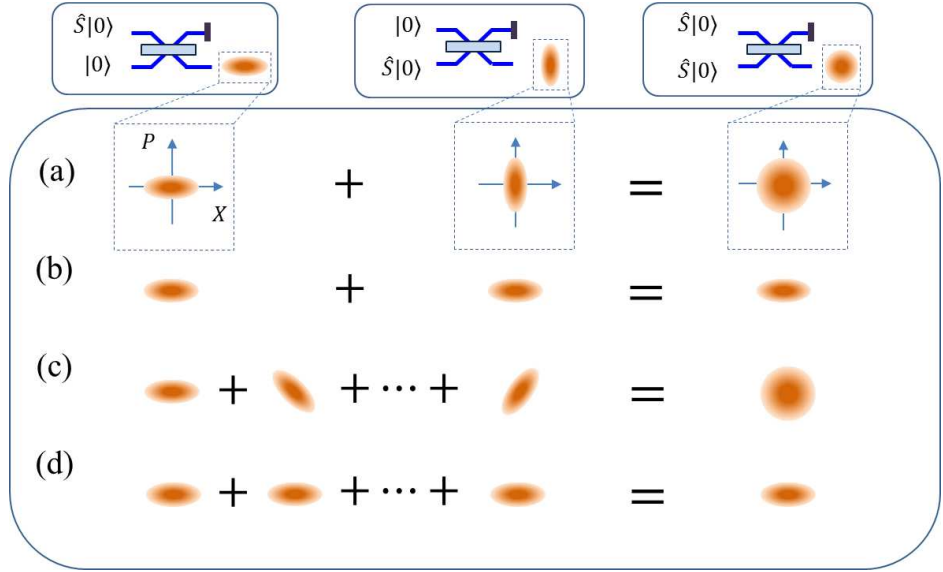} {}
\end{center}
\caption{Qualitative description of the output states in phase space. The output states correspond to (a) without and (b) with WFS in a two-port beam splitter, (c) without and (d) with WFS in a multiple-port disordered medium. Note that each state of the left-hand side of the "equality" represents the output light which corresponds to each individual input beam as input separately. The term of the right-hand side denotes the final output state which is related to the case of all the initial lights as input simultaneously. When mixing two output squeezed states (a) with different squeezing angles, the final output state would not achieve the optimal minimum variance, and (b) with the same squeezing angle, the final output state would achieve the optimal minimum variance. If mixing multiple squeezed states (c) with random squeezing angles, the final state would have a quantum fluctuation above the SNL, and (d) with the same squeezing angle, the final state would achieve the optimal minimum variance. It is obvious that rotating all squeezing angles in the same direction is a valid method of suppressing the final quantum noise. Actually, this reduction occurs as a result of destructive interference of quantum noise. Since WFS plays an essential role in manipulating the direction of squeezing angle, it can modulate the final quantum fluctuation. }
\label{illustration}
\end{figure*}

Consider the case II where the WFS is performed as depicted in Fig. \ref{bs12}(b). The WFS is actually equivalent to two phase shifters acting on two input modes. Without loss of generality, we may consider the two phase shifters to be $\phi_1$ and $\phi_2$, respectively. 

After $|\Psi^{\rm{in}}_3 \rangle$ propagates through the BS, the output state can be written as
\begin{align}
\label{eq22}
|\Psi^{\rm{out,w}}_3\rangle = e^{(r/2) [(\hat{a}_0^{\rm{out} })^2 - (\hat{a}_0^{\rm{out}\dagger})^{ 2}] - (r/2) [(\hat{a}_1^{\rm{out} })^2 - (\hat{a}_1^{\rm{out}\dagger})^{ 2}]}|0\rangle \otimes |0\rangle,
\end{align}
where we have set $\phi_1 = \pi/2$ and $\phi_2 = 0$ (see Append. \ref{bspresence}). From Eq. (\ref{eq22}), it is found that each output mode is a single-mode squeezed state. Then the variance of quadrature of output mode $0$ is calculated out to
\begin{align}
\label{minus2r}
\langle (\Delta \hat{x}_{0}^{\rm{out,w}})^2 \rangle = e^{-2r},
\end{align} 
where the output noise is below the SNL. This is similar to the single-mode squeezed states in the disordered medium with WFS.

Comparing Eqs. (\ref{plus2r}) and (\ref{minus2r}), one can easily see that the output noise with WFS is smaller than the one without WFS. This indicates that the WFS can reduce the variance of output quadrature, which is due to the destructive interference of quantum noise \cite{elste2009}. A detailed explanation will be presented in next subsection.

\subsection{Quantum-noise reduction resulting from uniform squeezing angles via WFS}

To illustrate quantum-noise reduction intuitively, we plot the output states in phase space in Fig. \ref{illustration}. Fig. \ref{illustration}(a) corresponds to the case I of a BS without WFS, the state of the left-hand side of the "equality" denotes the output beam which results from each single beam as input separately while the state on the right-hand side indicates the final output state with both beams as input simultaneously. Without WFS, the output states of the left-hand side present different squeezing angles. As a consequence, the final output state is the thermal state which shows a quantum fluctuation above the SNL. 

With the help of WFS in a BS, as depicted in Fig. \ref{illustration}(b), the output states of the left-hand side of the "equality" demonstrate in an uniform squeezing angle. As a result, the final output maintains a sub-shot noise. In other words, this final state has a quantum fluctuation below the SNL. We call this phenomenon the destructive interference of quantum noise.

By contrast, Figs. \ref{illustration}(c) and \ref{illustration}(d) are related to the disordered medium, a multiple-port optical device, without and with WFS. Similarly, in the absence of WFS [Fig. \ref{illustration}(c)], the output states of the left-hand side have randomly distributed squeezing angles. This leads to the final output state with a quantum fluctuation which is larger than the SNL. However, in the presence of WFS [Fig. \ref{illustration}(d)], squeezing angles are distributed uniformly. As a consequence, the final output state has a squeezed quantum fluctuation below the SNL.

From Fig. \ref{illustration}, it is evident that WFS modulates the quantum fluctuation via rotating the squeezing angles. Without WFS, the output squeezing angles are randomly distributed which gives rise to a final output quantum fluctuation above the SNL. In the presence of WFS, it rotates the squeezing angles in the same direction which makes the final output quantum fluctuation below the SNL owing to the destructive interference of quantum noise.

\section{Conclusion}
In summary, the effect of wavefront shaping on the quantum fluctuations of quadratures of scattered modes is investigated. It is clarified that wavefront shaping leads to quantum-noise reduction of scattered beams. Particularly, when the input is the single-mode squeezed states, the quantum fluctuation can always be decreased below the shot-noise level. If the two-mode squeezed states are considered as input, the quantum fluctuation can be degraded, but it is not always below the shot-noise level. As a matter of fact, there exits a threshold $g_{\star}$ which determines whether the reduced quantum noise reaches below the shot-noise level. When the input squeezing parameter is smaller than the threshold $g < g_{\star} \approx 1.06$, the reduced quantum noise can always achieve below the SNL whereas the one is always above the shot-noise level when $g > g_{\star} \approx 1.06$. Moreover, with the increasing of disorder strength, the degree of reduction of the quantum noise decreases. Above all, the quantum-noise reduction results from destructive interference of quantum noise via wavefront shaping. 

These results may have applications in quantum information processing, for instance, sub-wavelength imaging \cite{putten2011,park2014,jang2018,chen2018}, where the disordered medium plays a role in focusing lights as a scattering superlens. Recall that our proposal with squeezed-state sources provides the output with a sub-shot noise, which may improve the resolution in imaging by boosting the signal-to-noise ratio.

\section{Acknowledge}
We thank Prof. Song Sun for his insightful suggestions. This work was supported by the Science Challenge Program (Grant No. TZ2018003-3) and National Natural Science Foundation of China (Grant Nos. 61875178 and 11605166).

\appendix
\subsection{Derivation of the summation of transmission and reflection coefficients of the scattered mode $b$}
\label{derivation}
The input-output relation of a disordered medium is given by
\begin{align}
\hat{a}_{b}^{\dagger} &= \sum_{a'}{t_{a'b}^{\ast} \hat{a}_{a'}^{\rm{in}\dagger} } +\sum_{b'} {r_{b'b}^{\ast} \hat{a}_{b'}^{\rm{in} \dagger}}, \\ \nonumber
\hat{a}_{b} &= \sum_{a'}{t_{a'b} \hat{a}_{a'}^{\rm{in}} } +\sum_{b'} {r_{b'b} \hat{a}_{b'}^{\rm{in}}},
\end{align}
where $t_{a'b}^{\ast}$ ($r_{b'b}^{\ast}$) is the conjugate of $t_{a'b}$ ($r_{b'b}$). According to the commutation relation $[\hat{a}_b, \hat{a}_b^{\dagger}] = 1$, one can easily obtain 
\begin{align}
\sum_{a'} |t_{a'b}|^2 + \sum_{b'} |r_{b'b}|^2 = 1,
\end{align}
where $[\hat{a}_{i}^{\rm{in}}, \hat{a}_{j}^{\rm{in}\dagger}] = \delta_{ij}$ ($i,j=a',b'$) has been used. Therefore, $\sum_{a'} T_{a'b} + \sum_{b'} R_{b'b} = 1$.

\subsection{Quantum variance in the absence of WFS}
\subsubsection{Single-mode squeezed states as input}
\label{appsinglemode}
The variance of $\hat{x}_b$ without WFS is given by
\begin{align}
\label{v2b}
\langle (\Delta \hat{x}_b)^2\rangle = \langle \hat{x}_b^2\rangle - \langle \hat{x}_b\rangle^2.
\end{align}
To obtain the variance $\langle (\Delta \hat{x}_b)^2\rangle$, it is necessary to calculate $ \langle \hat{x}_b\rangle$ and $\langle \hat{x}_b^2\rangle$.

In the absence of WFS, the mean value of $\hat{x}_b$ in Eq. (\ref{x0}) is found to be
\begin{align}
\langle \hat{x}_b \rangle =& \sum_{a'} {\sqrt{T_{a'b}} [ \cos \phi_{a'b} \langle \hat{x}_{a'}^{\rm{in}} \rangle - \sin \phi_{a'b} \langle \hat{p}_{a'}^{\rm{in}} \rangle] }.
\end{align}
The expectation value of $\hat{x}_b^2$ is worked out as
\begin{align}
\langle \hat{x}_b^2 \rangle =& \sum_{a'} {T_{a'b} [ \cos^2 \phi_{a'b} \langle (\hat{x}_{a'}^{\rm{in}})^2 \rangle + \sin^2 \phi_{a'b} \langle (\hat{p}_{a'}^{\rm{in}})^2 \rangle - \cos \phi_{a'b} \sin \phi_{a'b} (\langle \hat{x}_{a'}^{\rm{in}} \hat{p}_{a'}^{\rm{in}} \rangle + \langle \hat{p}_{a'}^{\rm{in}} \hat{x}_{a'}^{\rm{in}} \rangle) ] }\\ \nonumber
& +  \sum_{b'} {R_{b'b} [ \cos^2 \phi_{b'b} \langle (\hat{x}_{b'}^{\rm{in}})^2 \rangle + \sin^2 \phi_{b'b} \langle (\hat{p}_{b'}^{\rm{in}})^2 \rangle - \cos \phi_{b'b} \sin \phi_{b'b} (\langle \hat{x}_{b'}^{\rm{in}} \hat{p}_{b'}^{\rm{in}} \rangle + \langle \hat{p}_{b'}^{\rm{in}} \hat{x}_{b'}^{\rm{in}} \rangle) ] }\\ \nonumber
& + \sum_{a'a''} \{ \sqrt{T_{a'b} T_{a''b}} [\cos \phi_{a' b} \cos \phi_{a''b} (\langle \hat{x}_{a'}^{\rm{in}} \hat{x}_{a''}^{\rm{in}} \rangle + \langle \hat{x}_{a''}^{\rm{in}} \hat{x}_{a'}^{\rm{in}} \rangle) + \sin \phi_{a' b} \sin \phi_{a''b} (\langle \hat{p}_{a'}^{\rm{in}} \hat{p}_{a''}^{\rm{in}} \rangle + \langle \hat{p}_{a''}^{\rm{in}} \hat{p}_{a'}^{\rm{in}} \rangle) \\ \nonumber
& - \cos \phi_{a' b} \sin \phi_{a''b} (\langle \hat{x}_{a'}^{\rm{in}} \hat{p}_{a''}^{\rm{in}} \rangle + \langle \hat{x}_{a''}^{\rm{in}} \hat{p}_{a'}^{\rm{in}} \rangle) ]  \}.
\end{align}
Based on Eq. (\ref{v2b}), the variance is then given by
\begin{align}
\langle (\Delta \hat{x}_{b})^2 \rangle =& 
\sum_{a'}{T_{a'b}[\cos^2 \phi_{a'b}\langle (\Delta \hat{x}_{a'}^{\rm{in}})^2 \rangle + \sin^2 \phi_{a'b}\langle (\Delta \hat{p}_{a'}^{\rm{in}})^2 \rangle - 2\cos \phi_{a'b} \sin \phi_{a'b} {\rm{cov}}(\hat{x}_{a'}^{\rm{in}}, \hat{p}_{a'}^{\rm{in}}))]} \\ \nonumber
& + \sum_{b'}{R_{b'b}[\cos^2 \phi_{b'b}\langle (\Delta \hat{x}_{b'}^{\rm{in}})^2 \rangle + \sin^2 \phi_{b'b}\langle (\Delta \hat{p}_{b'}^{\rm{in}})^2 \rangle - 2\cos \phi_{b'b} \sin \phi_{b'b} {\rm{cov}}(\hat{x}_{b'}^{\rm{in}}, \hat{p}_{b'}^{\rm{in}}) ]}\\ \nonumber
& + \sum_{a'\neq a''} \{ \sqrt{T_{a'b} T_{a''b}} [2\cos \phi_{a' b} \cos \phi_{a''b}  \langle \Delta^2(\hat{x}_{a'}^{\rm{in}} \hat{x}_{a''}^{\rm{in}}) \rangle + 2\sin \phi_{a' b} \sin \phi_{a''b}  {\rm{cov}}(\hat{p}_{a'}^{\rm{in}}, \hat{x}_{a''}^{\rm{in}}) \\ \nonumber
&-2\cos \phi_{a'b} \sin \phi_{a''b} {\rm{cov}}(\hat{x}_{a'}^{\rm{in}}, \hat{p}_{a''}^{\rm{in}})] \},
\end{align}
where the covariance fuction is defined as ${\rm{cov}}(\hat{Y},\hat{Z}) \equiv \frac{1}{2} (\langle \hat{Y} \hat{Z} \rangle +\langle \hat{Z} \hat{Y} \rangle) - \langle \hat{Y}\rangle \langle \hat{Z} \rangle $. By averaging over all disorder ensembles, the variance is found to be
\begin{align}
\label{sqz1var200}
\overline{\langle (\Delta \hat{x}_{b})^2 \rangle} = \sum_{a'} \overline{T_{a'b}} [\frac{1}{2} \langle (\Delta \hat{x}_{a'}^{\rm{in}})^2 \rangle + \frac{1}{2} \langle (\Delta \hat{p}_{a'}^{\rm{in}})^2 \rangle] + \sum_{b'} \overline{R_{b'b}},
\end{align}
where we have used $\overline{\cos^2 \phi_{a'b}} = \overline{\sin^2 \phi_{a'b}} = \frac{1}{2}$ and $\overline{\cos \phi_{a'b}\sin \phi_{a'b}} = \overline{\sin \phi_{a'b}\sin \phi_{a''b}} = \overline{\cos \phi_{a'b}\cos \phi_{a''b}} =0$.  The variance can be reduced to
\begin{align}
\label{varx2app}
\overline{\langle (\Delta \hat{x}_{b})^2 \rangle} = 1+ \frac{K}{Ns} [\cosh (2r) - 1],
\end{align}
where we have used $\langle (\Delta \hat{x}_{a'}^{\rm{in}})^2 \rangle = e^{-2r}$, $\langle (\Delta \hat{p}_{a'}^{\rm{in}})^2 \rangle = e^{2r}$, $\sum_{a'} T_{a'b} + \sum_{b'} R_{b'b} = 1$, $\overline{T_{a'b}} = 1/(Ns)$, and $K$ is the number of input modes. When $K = 1$, namely one single-mode squeezed state as input, Eq. (\ref{varx2app}) is reduced to $\overline{\langle (\Delta \hat{x}_{b})^2 \rangle} = 1+  [\cosh (2r) - 1]/ (Ns)$ which reproduces the result in Ref. \cite{lodahl2006b}. According to Eq. (\ref{varx2app}), it is clear that the averaged variance is always greater than one when $r>0$. This result indicates that the averaged quantum fluctuation of quadrature is always above the shot-noise level. Similarly, one can obtain that 
\begin{align}
\label{varp2app}
\overline{\langle (\Delta \hat{p}_{b})^2 \rangle} = 1+ \frac{K}{Ns} [\cosh (2r) - 1].
\end{align}

\subsubsection{Two-mode squeezed states as input}
\label{apptwomode}
Consider two-mode squeezed states as input, in the absence of WFS, the expectation value of $\hat{x}_b$ is found to be
\begin{align}
\label{sqz2x}
\langle \hat{x}_b \rangle =& \sum_{a'} {\sqrt{T_{a'b}} [ \cos \phi_{a'b} \langle \hat{x}_{a'}^{\rm{in}} \rangle - \sin \phi_{a'b} \langle \hat{p}_{a'}^{\rm{in}} \rangle] }.
\end{align}
The mean value of $\hat{x}_b^2$ is recast as
\begin{align}
\label{sqz2x2}
\langle \hat{x}_b^2 \rangle =& \sum_{a'} {T_{a'b} [ \cos^2 \phi_{a'b} \langle (\hat{x}_{a'}^{\rm{in}})^2 \rangle + \sin^2 \phi_{a'b} \langle (\hat{p}_{a'}^{\rm{in}})^2 \rangle - \cos \phi_{a'b} \sin \phi_{a'b} (\langle \hat{x}_{a'}^{\rm{in}} \hat{p}_{a'}^{\rm{in}} \rangle + \langle \hat{p}_{a'}^{\rm{in}} \hat{x}_{a'}^{\rm{in}} \rangle) ] }\\ \nonumber
& +  \sum_{b'} {R_{b'b} [ \cos^2 \phi_{b'b} \langle (\hat{x}_{b'}^{\rm{in}})^2 \rangle + \sin^2 \phi_{b'b} \langle (\hat{p}_{b'}^{\rm{in}})^2 \rangle - \cos \phi_{b'b} \sin \phi_{b'b} (\langle \hat{x}_{b'}^{\rm{in}} \hat{p}_{b'}^{\rm{in}} \rangle + \langle \hat{p}_{b'}^{\rm{in}} \hat{x}_{b'}^{\rm{in}} \rangle) ] }\\ \nonumber
& + \sum_{a'a''} \{ \sqrt{T_{a'b} T_{a''b}} [\cos \phi_{a' b} \cos \phi_{a''b} (\langle \hat{x}_{a'}^{\rm{in}} \hat{x}_{a''}^{\rm{in}} \rangle + \langle \hat{x}_{a''}^{\rm{in}} \hat{x}_{a'}^{\rm{in}} \rangle) + \sin \phi_{a' b} \sin \phi_{a''b} (\langle \hat{p}_{a'}^{\rm{in}} \hat{p}_{a''}^{\rm{in}} \rangle + \langle \hat{p}_{a''}^{\rm{in}} \hat{p}_{a'}^{\rm{in}} \rangle) \\ \nonumber
& - \cos \phi_{a' b} \sin \phi_{a''b} (\langle \hat{x}_{a'}^{\rm{in}} \hat{p}_{a''}^{\rm{in}} \rangle + \langle \hat{x}_{a''}^{\rm{in}} \hat{p}_{a'}^{\rm{in}} \rangle) ]  \}.
\end{align}
Combining Eqs. (\ref{v2b}), (\ref{sqz2x}), and (\ref{sqz2x2}), one can obtain the variance
\begin{align}
\langle (\Delta \hat{x}_{b})^2 \rangle =& 
\sum_{a'}{T_{a'b}[\cos^2 \phi_{a'b}\langle (\Delta \hat{x}_{a'}^{\rm{in}})^2 \rangle + \sin^2 \phi_{a'b}\langle (\Delta \hat{p}_{a'}^{\rm{in}})^2 \rangle - 2\cos \phi_{a'b} \sin \phi_{a'b} {\rm{cov}}(\hat{x}_{a'}^{\rm{in}}, \hat{p}_{a'}^{\rm{in}})]} \\ \nonumber
& + \sum_{b'}{R_{b'b}[\cos^2 \phi_{b'b}\langle (\Delta \hat{x}_{b'}^{\rm{in}})^2 \rangle + \sin^2 \phi_{b'b}\langle (\Delta \hat{p}_{b'}^{\rm{in}})^2 \rangle - 2\cos \phi_{b'b} \sin \phi_{b'b} {\rm{cov}}(\hat{x}_{b'}^{\rm{in}}, \hat{p}_{b'}^{\rm{in}}) ]}\\ \nonumber
& + \sum_{a'\neq a''} \{ \sqrt{T_{a'b} T_{a''b}} [2\cos \phi_{a' b} \cos \phi_{a''b}  \langle \Delta^2(\hat{x}_{a'}^{\rm{in}} \hat{x}_{a''}^{\rm{in}}) \rangle + 2\sin \phi_{a' b} \sin \phi_{a''b}  {\rm{cov}}(\hat{p}_{a'}^{\rm{in}} \hat{x}_{a''}^{\rm{in}}) \rangle \\ \nonumber
&-2\cos \phi_{a'b} \sin \phi_{a''b} {\rm{cov}}(\hat{x}_{a'}^{\rm{in}}, \hat{p}_{a''}^{\rm{in}}) ] \}.
\end{align}
By averaging over all disorder ensembles, the variance is found to be
\begin{align}
\overline{\langle (\Delta \hat{x}_{b})^2 \rangle} = \sum_{a'} \overline{T_{a'b}} [\frac{1}{2} \langle (\Delta \hat{x}_{a'}^{\rm{in}})^2 \rangle + \frac{1}{2} \langle (\Delta \hat{p}_{a'}^{\rm{in}})^2 \rangle] + \sum_{b'} \overline{R_{b'b}},
\end{align}
where we have used $\overline{\cos^2 \phi_{a'b}} = \overline{\sin^2 \phi_{a'b}} = \frac{1}{2}$ and $\overline{\cos \phi_{a'b}\sin \phi_{a'b}} = \overline{\sin \phi_{a'b}\sin \phi_{a''b}} = \overline{\cos \phi_{a'b}\cos \phi_{a''b}} =0$.

The mean variance can be simplified to
\begin{align}
\label{varx3}
\overline{\langle (\Delta \hat{x}_{b})^2 \rangle} = 1+ \frac{2 K}{Ns}  \sinh^2 g,
\end{align}
where we have utilized $\langle (\Delta \hat{x}_{a'}^{\rm{in}})^2 \rangle = \langle (\Delta \hat{p}_{a'}^{\rm{in}})^2 \rangle = 2 \sinh^2 g + 1$, $\sum_{a'} T_{a'b} + \sum_{b'} R_{b'b} = 1$, and $\overline{T_{a'b}} = 1/(Ns)$. Similarly, it is found that
\begin{align}
\label{varp3}
\overline{\langle (\Delta \hat{p}_{b})^2 \rangle} = 1+ \frac{2 K}{Ns} \sinh^2 g.
\end{align}

\subsection{Rayleigh distribution}
\label{rayleigh}
According to Refs. \cite{goodman2015,starshynov2016}, $\sqrt{T_{a'b}}$ obeys Rayleigh distribution $P(\sqrt{T_{a'b}})$, where $P(y) = \frac{y}{\sigma^2} e^{-y^2/(2 \sigma^2)}$ with $\sigma$ being a constant parameter of the distribution. For a variable $\theta$ of Rayleigh distribution, one can obtain the averages 
\begin{align}
\label{ray1}
\overline{\theta} = \int_0^{\infty} \theta P(\theta) d \theta,
\end{align}
\begin{align}
\label{ray2}
\overline{\theta^2} = \int_0^{\infty} \theta^2 P(\theta) d \theta.
\end{align}
Comparing Eqs. (\ref{ray1}) and (\ref{ray2}), we can find that \begin{align}
\label{ray3}
 \frac{\pi}{4}\overline{\theta^2} = \overline{\theta}^2 = \overline{\theta_1}\overline{\theta_2},
\end{align}
where $\theta_1$ and $\theta_2$ are two independent random variables of Rayleigh distribution $P(y) = \frac{y}{\sigma^2} e^{-y^2/(2 \sigma^2)}$. Let $\theta = \sqrt{T_{a'b}}$, $\theta_1 = \sqrt{T_{a''b}}$, and $\theta_2 = \sqrt{T_{a'+K/2, b}}$, then Eq. (\ref{ray3}) can be rewritten as
\begin{align}
 \frac{\pi}{4}\overline{T_{a'b}}  =  \overline{\sqrt{T_{a''b}}}\overline{\sqrt{T_{a'+K/2, b}}} =  \overline{\sqrt{T_{a''b}}\sqrt{T_{a'+K/2, b}}}.
\end{align}
Hence, it is obtained that
\begin{align}
\overline{\sqrt{T_{a''b}T_{a'+K/2, b}}} = \frac{\pi}{4} \overline{T_{a'b}}.
\end{align}
The prove is complete.

\subsection{Quadrature variance of output state after a BS}

we consider the situation where single-mode squeezed vacuum states are mixed by a BS. The input state is given by
\begin{align}
\label{inputstate0}
|\Psi^{\rm{in}}_3\rangle &= \hat{S}_{a_0}(r) |0\rangle \otimes \hat{S}_{a_1}(r) |0\rangle,
\end{align}
where $\hat{S}_{a_0}(r) = e^{(r/2) [(\hat{a}_0^{\rm{in}\dagger})^{ 2}-  (\hat{a}_0^{\rm{in} })^2] }$ and $\hat{S}_{a_1}(r) = e^{(r/2) [(\hat{a}_1^{\rm{in}\dagger})^{ 2}-  (\hat{a}_1^{\rm{in} })^2] }$.

\subsubsection{In the absence of WFS}
\label{bsabsence}
As depicted in Fig. \ref{bs12}(a), without WFS the output state after BS can be described as
\begin{align}
\left(
\begin{array}
[c]{c}%
\hat{a}_{0}^{\rm{out}} \\
\hat{a}_{1}^{\rm{out}}%
\end{array}
\right)
=
\hat{T}_{\rm{BS}}
\left(
\begin{array}
[c]{c}%
\hat{a}_{0}^{\rm{in}} \\
\hat{a}_{1}^{\rm{in}}%
\end{array}
\right)
,
\end{align}
where
\begin{align}
\hat{T}_{\rm{BS}}  =\frac{1}{\sqrt{2}}\left(
\begin{array}
[c]{cc}%
1 & i\\
i & 1%
\end{array}
\right)
.\label{bs}
\end{align}
The input modes can be expressed in terms with the output states as
\begin{align}
\left(
\begin{array}
[c]{c}%
\hat{a}_{0}^{\rm{in}} \\
\hat{a}_{1}^{\rm{in}}%
\end{array}
\right)
=
\hat{T}_{\rm{BS}}^{-1}
\left(
\begin{array}
[c]{c}%
\hat{a}_{0}^{\rm{out}} \\
\hat{a}_{1}^{\rm{out}}%
\end{array}
\right)
.\label{bs0x}
\end{align}
Inserting Eq. (\ref{bs0x}) into Eq. (\ref{inputstate0}), the output state can be written as
\begin{align}
|\Psi^{\rm{out}}_3\rangle &= e^{i r [\hat{a}_0^{\rm{out}\dagger} \hat{a}_1^{\rm{out}\dagger} +  \hat{a}_0^{\rm{out} } \hat{a}_1^{\rm{out} }]}|0\rangle \otimes |0\rangle.
\end{align}
The variance of the output mode $a_0$ is then worked out as
\begin{align}
\label{tmsq}
\langle (\Delta \hat{x}_{0}^{\rm{out}})^2 \rangle = 2 \sinh^2 r + 1,
\end{align} 
where $\hat{x}_{0}^{\rm{out}} \equiv \hat{a}_{0}^{\rm{out} \dagger} + \hat{a}_{0}^{\rm{out}}$.

\subsubsection{In the presence of WFS}
\label{bspresence}
In the presence of WFS, the output state after BS is found to be
\begin{align}
\left(
\begin{array}
[c]{c}%
\hat{a}_{0}^{\rm{out}} \\
\hat{a}_{1}^{\rm{out}}%
\end{array}
\right)
=
\hat{T}_{\rm{BS}} \hat{T}_{{\phi}}
\left(
\begin{array}
[c]{c}%
\hat{a}_{0}^{\rm{in}} \\
\hat{a}_{1}^{\rm{in}}%
\end{array}
\right)
,
\end{align}
where
\begin{align}
\hat{T}_{\phi}  =\left(
\begin{array}
[c]{cc}%
e^{i \phi_1} & 0\\
0 & e^{i \phi_2}%
\end{array}
\right)
,\label{bs2}
\end{align} 
\begin{align}
\hat{T}_{\rm{BS}}  =\frac{1}{\sqrt{2}}\left(
\begin{array}
[c]{cc}%
1 & i\\
i & 1%
\end{array}
\right)
.\label{bs}
\end{align}
The operators of input modes can be expressed as
\begin{align}
\left(
\begin{array}
[c]{c}%
\hat{a}_{0}^{\rm{in}} \\
\hat{a}_{1}^{\rm{in}}%
\end{array}
\right)
=
(\hat{T}_{\rm{BS}} \hat{T}_{{\phi}})^{-1}
\left(
\begin{array}
[c]{c}%
\hat{a}_{0}^{\rm{out}} \\
\hat{a}_{1}^{\rm{out}}%
\end{array}
\right)
.
\end{align}
Then the output state is given by
\begin{align}
|\Psi^{\rm{out,w}}_3 \rangle =& \exp\{\frac{r}{4}[(e^{i 2 \phi_1} - 1)(\hat{a}_{0}^{\rm{out} \dagger})^2 + (1 - e^{-i 2 \phi_1})(\hat{a}_{0}^{\rm{out}})^2  \\ \nonumber
&+ (1 - e^{i 2 \phi_1})(\hat{a}_{1}^{\rm{out} \dagger})^2 + (e^{-i 2 \phi_1} - 1)(\hat{a}_{1}^{\rm{out}})^2 \\ \nonumber &+ i(2 + 2 e^{i2\phi_1}) \hat{a}_{0}^{\rm{out} \dagger}\hat{a}_{1}^{\rm{out} \dagger}  + i(2 + 2 e^{-i2\phi_1}) \hat{a}_{0}^{\rm{out} }\hat{a}_{1}^{\rm{out}} ]\} |0\rangle \otimes |0\rangle,
\label{phieq}
\end{align}
where we have set $\phi_2 = 0$ as a reference. When $\phi_1 = \pi/2$, Eq. (\ref{phieq}) is simplified to
\begin{align}
|\Psi^{\rm{out,w}}_3\rangle = e^{(r/2) [(\hat{a}_0^{\rm{out} })^2 - (\hat{a}_0^{\rm{out}\dagger})^{ 2}] - (r/2) [(\hat{a}_1^{\rm{out} })^2 - (\hat{a}_1^{\rm{out}\dagger})^{ 2}]}|0\rangle \otimes |0\rangle,
\end{align}
where each output mode is a single-mode squeezed state. The variance of the output mode $0$ is then worked out as
\begin{align}
\langle (\Delta \hat{x}_{0}^{\rm{out,w}})^2 \rangle = e^{-2r}.
\end{align}

\end{document}